\documentclass[notitlepage,aps,prd,amsmath,floats,floatfix,twocolumn,superscriptaddress,nofootinbib,showpacs]{revtex4}

\usepackage{amsfonts,amsmath,wasysym,epsfig,graphicx,verbatim,color,subfigure,graphicx}
\graphicspath{{Images/}}

\usepackage[range-phrase=--,range-units=single,retain-unity-mantissa=false]{siunitx}
\usepackage{amsmath}
\usepackage{amssymb}
\usepackage{amsfonts}
\usepackage{wasysym}
\usepackage{bm}
\usepackage{color}
\usepackage[normalem]{ulem}
\usepackage{xcolor}
\usepackage{color,soul}

\usepackage{marginnote}
\usepackage{braket}
\usepackage[none]{hyphenat}
\usepackage{bm}
\usepackage{comment}
\usepackage{multirow}
\usepackage{xspace}
\usepackage[toc,page]{appendix}
\usepackage{float} 
\usepackage{hyperref}
\usepackage{booktabs}
\newcommand{\asi}{a-Si\xspace}

\newcommand{\hafnia}{HfO$_{2}$\xspace}
\newcommand{\silica}{SiO$_{2}$\xspace}
\newcommand{\tantala}{Ta$_{2}$O$_{5}$\xspace}

\newcommand{\titania}{TiO$_{2}$\xspace}

\newcommand{\GA}{GaAs\xspace}
\newcommand{\AG}{AlGaAs\xspace}
\newcommand{\GAG}{GaAs/AlGaAs\xspace}
\newcommand{\AGfull}{Al$_{0.92}$Ga$_{0.08}$As\xspace}
\newcommand{\GAGfull}{GaAs/Al$_{0.92}$Ga$_{0.08}$As\xspace}

\ifdefined\COMMENTS

\else

\fi

\newcommand{\SU}{\affiliation{Department of Physics, Syracuse University, Syracuse, New York 13244, USA}}
\newcommand{\Cardiff}{\affiliation{School of Physics and Astronomy, Cardiff University, Cardiff CF24 3AA, United Kingdom}}
\newcommand{\HWS}{\affiliation{Department of Physics, Hobart William Smith Colleges, Geneva, New York 14456, USA}}
\newcommand{\LHO}{\affiliation{LIGO Hanford Observatory, Richland, Washington 99352, USA}}

\newcommand{\UA}{\affiliation{Wyant College of Optical Sciences, University of Arizona, Tucson, Arizona, 85721 USA}}

\date{\today}

\begin{document}

\title{Cryogenic Mechanical Loss of \GAG Crystalline Coatings}

\author{Nicholas A. Didio}\SU
\author{Steven D. Penn}\SU\HWS\email{sdpenn@syr.edu}
\author{Satoshi Tanioka}\Cardiff
\author{Elenna M. Capote}\LHO
\author{Garrett D. Cole}\UA
\author{Stefan W. Ballmer}\SU

\begin{abstract}
Optical coatings with minimal thermal noise are essential for high-precision, laser-based  metrology experiments. Substrate-transferred crystalline 
\GAGfull coatings  have the lowest 
thermal noise for high-reflectivity mirror coatings with $\lambda =900$ nm -- 12 $\mu$m. 
Time-frequency stabilization experiments have demonstrated that \GAG mirror coatings have a lower noise floor than amorphous mirror coatings.  
These mirrors are now being developed at sizes $\ge 300$~mm for gravitational-wave detectors. In order to understand the coating's fundamental loss mechanisms, which cause thermal noise, we have performed  the first measurement of the mechanical loss at cryogenic temperatures using a cryogenic, multimodal, gentle nodal suspension system, which operates from 12 K to room temperature.  This initial measurement used a silicon substrate whose asymmetry produced excess friction with the nodal support.  Accounting for this friction and Akhiezer loss in the substrate and coating, the mechanical loss agrees well with the thermal noise observed in experiments using cryogenic fixed-cavities with \GAG mirrors.


\end{abstract}

\maketitle

\section{\label{sec:Intro} Introduction}
 Thermal noise in mirror coatings is a limiting noise source in very high precision optical measurements, such as stabilized time-frequency experiments~\cite{Lee2026PRL,Kessler2012, Matei2017} and interferometric gravitational wave detectors~\cite{Punturo2010, Harry2012optical, Aso2013, Aasi2015, Adhikari2020, Evans2021horizon}.

Brownian thermal noise is observed in systems in thermal energy equilibrium where dissipation mechanisms cause nonresonant motion. The Fluctuation-Dissipation Theorem (FDT) of Callen and Welton~\cite{Callen1951irreversibility} predicts the power spectral density of the thermal noise as a function of the thermal energy and the dissipation, characterized by the energy loss per cycle or loss angle, $\phi = \Delta E/\left(2 \pi E\right)$.

The current generation of advanced gravitational wave detectors (GWDs), are
sensitivity limited in their observation band by the thermal noise of the high-reflectivity (HR) coatings on their test mass mirrors\cite{Gras2018direct, Buikema2020sensitivity, O4_sensitivity}.  The rate of quantum noise reduction through frequency-dependent squeezing and increased laser power continues to increasingly exposed the sensitivity limitation of the test mass coating thermal noise (CTN)\cite{O4_sensitivity}.

The power spectral density of the CTN, $S_{\mathrm{CTN}}(f)$, is described as \cite{Harry2002thermal, Penn2019mechanical}
\begin{align}
    S_x(f) &= 2k_{\mathrm{B}}T\frac{(1-\sigma^2)}{\pi^{3/2}wYf}\phi_{\mathrm{eff}},
    \label{eq.PSDctn}
\end{align}
where $k_{\mathrm{B}}$, $T$, $f$,  $w$, $\sigma$, $Y$, and $\phi_{\mathrm{eff}}$ are, respectively, the Boltzmann constant, the temperature, the frequency, the beam radius, the Poisson's ratio of the substrate, the Young's modulus of the substrate,  and the effective mechanical loss angle.
This effective loss is given by the relationship between the losses of the substrate and coating. If one makes the simplifying, but usually unphysical,  assumption that the coating material has a uniform, isotropic elasticity characterized by a single loss angle $\phi_{\mathrm{c}}$, then \cite{Harry2002thermal, Penn2019mechanical}
\begin{align}\label{eq.PhiEffective}
    \phi_{\mathrm{eff}} = \phi_{\mathrm{s}} + \phi_{\mathrm{c}}\frac{2d - 4d\sigma}{\sqrt{\pi}w(1-\sigma)}
\end{align}
where $\phi_{\mathrm{s}}$, $\phi_{\mathrm{c}}$, and $d$, are the loss angles of the substrate and coating, and the coating thickness, respectively.

The \GAG mirror coating is a single crystal composed of multiple alternating layers of \GA and \AGfull. \GA and \AG crystals have a zinc-blende structure that is described by 3 complex elastic constants. The ratio of the imaginary-to-real components of the elastic constants are the 3 loss angles that define the thermal noise.  In order to independently determine the 3 loss angles, one would need to measure the elastic loss for a range of modes with independent energy fractions (dilution factors) for the 3 elastic constants.
In our measurements, the uncertainty of the loss was too high to reliably determine all three loss angles.  Therefore, we use a single loss angle model that is an effective average over the three.

As seen in Eqn.~(\ref{eq.PSDctn}) the CTN may be reduced, and the detectors sensitivity improved, by developing coatings with lower mechanical loss. In order to understand, and potentially reduce, a material's dissipation, one can measure the loss angle as a function of temperature.  The spectrum will display Debye loss peaks at the activation energy of an Arrhenius relaxation process.~\cite{Debye1912Isolatoren}.  This initial measurement did not fulfill this goal because the loss mechanisms in the silicon wafer substrate dominated the loss at all temperatures above 25 K. 

Eqn.~(\ref{eq.PSDctn}) also shows that the CTN may be lowered by reducing temperature as long as the mechanical loss remains constant. Thus measurements of the cryogenic elastic loss for \GAG coatings should inform whether cryogenic GW detectors~\cite{Punturo2010, Adhikari2020, KAGRA2020} would gain sensitivity by adopting \GAG mirror coatings.

The mirror coatings in current GW detectors are ion-beam sputtered, amorphous, metal-oxide, dielectric Bragg reflectors. The Initial LIGO~\cite{Abbott:2007kv} coatings of \tantala/\silica had a CTN~\cite{Harry2002thermal} that was well above the requirements for Advanced LIGO~\cite{aLIGO2015}. The quieter \titania-doped \tantala/\silica coatings developed in 2007 for Advanced LIGO~\cite{Harry_2007} have now been used to detect over 390 GW events~\cite{GWTC5Intro}.  However, over the following 18 years, despite a significant research effort, the CTN in amorphous mirror coatings has been reduced by only $\approx 26$\%~\cite{fazio2025lowthermalnoisemirror}.

\GAG coatings are widely used in precision frequency experiments and are being developed in large diameters ($\ge 300$ mm) for GW detectors.\cite{Cole2013, Chalermsongsak2016, Marchio2018, Penn2019mechanical, Tanioka2023}.  
\GAG coatings  have a room temperature  mechanical loss of $\phi_{\mathrm{c}} \lesssim 2.5 \times 10^{-5}$~\cite{Cole2013}, with a  measured CTN that is 5--10$\times$ lower than the Advanced LIGO coatings~\cite{Gras2018direct}.
In a cryogenic optomechanical cantilever experiment~\cite{Cole2012}, \GAG coatings show a thermal noise consistent with a mechanical loss of $\phi_{\mathrm{c}}\approx4.5\times10^{-6}$.

\begin{figure}[htbp]
    \centering
    \graphicspath{ {./Figures/} }
    \includegraphics[width=8.6cm]{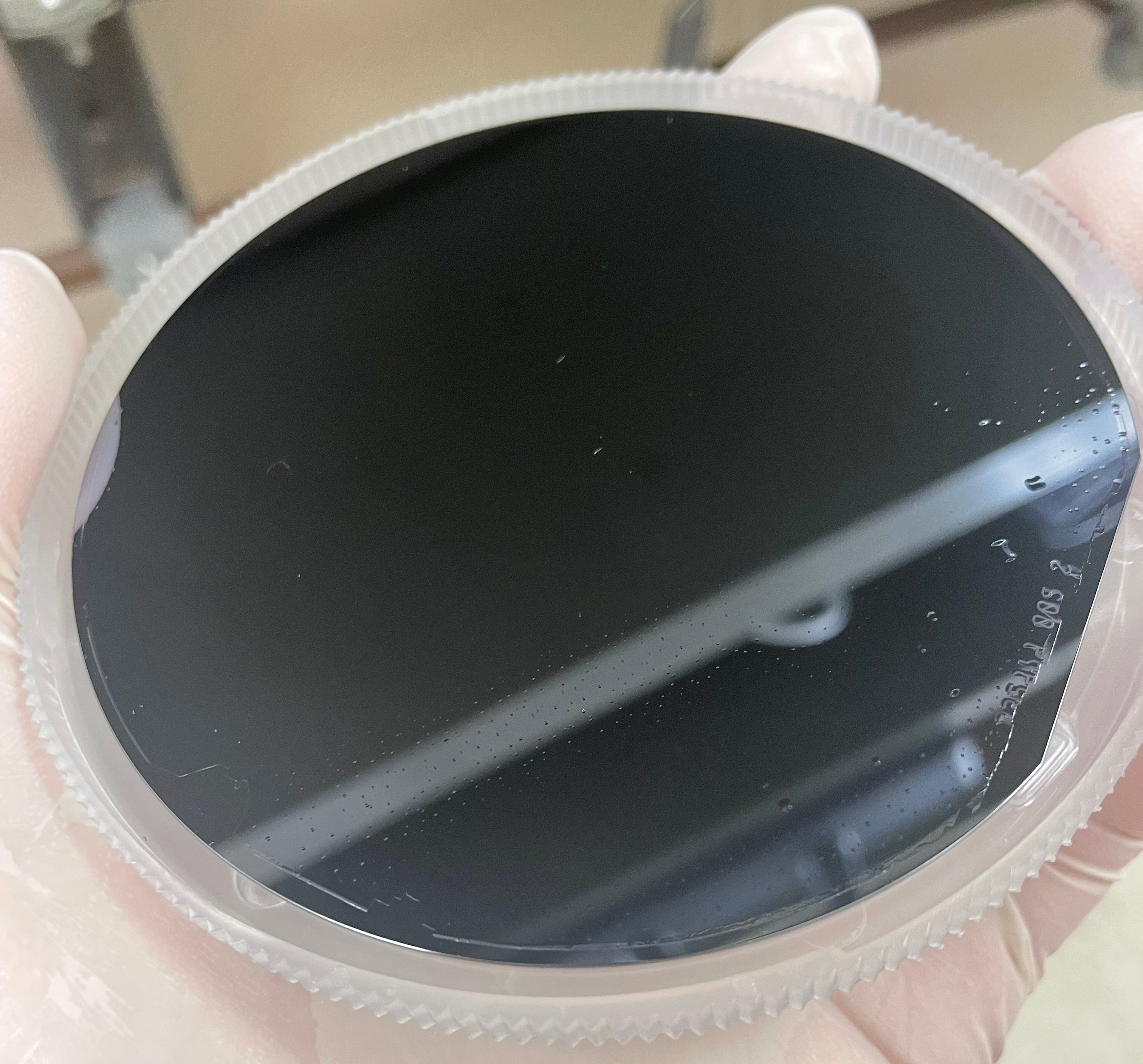}
    \caption{The \GAG-coated silicon disk. The flat on the silicon substrate indicates the $(110)$ plane. The \GAG $[0\bar{1}1]$  orientation was aligned to this flat. The white streak across the sample is simply a reflection of the wall opposite the camera.}
    \label{fig:AlGaAsCoating}
\end{figure}

In this work, we present mechanical loss measurements of a \GAG coating measured using a multimodal, cryogenic, gentle nodal suspension~\cite{Cesarini2009Gentle,  Vajente2017GeNS} (Cryo-GeNS) system for frequencies $f=390-8100\, \unit{Hz}$ and temperatures $T=12-300\,\unit{K}$. The silicon wafer substrate\cite{Ultrasil} (100\,\unit{mm} \invdiameter, $0.5\,\unit{mm}$ thick) had a flat (chord cut) to denote the (110) axis.  The \GAG coating had high-reflectivity coating ($R=0.99999,\,\lambda=1064$~\unit{nm}) and was designed to minimize thermo-optic noise\cite{Chalermsongsak2016}.  Thermo-optic noise is frequency noise in the reflected beam from the correlated thermoelastic and thermorefractive fluctuations.  Optimization involves adjusting the coating layer thickness in order to cancel, to first order, the thermoelastic and thermorefractive effects while maintaining high-reflectivity\cite{evans_2008}.  The coating was bonded with a flat such that the [0$\bar{1}$1] axis aligned with the substrate flat. (See Fig. \ref{fig:AlGaAsCoating}).  This sample had a smattering of small blisters that arose on the silicon-coating interface. While not optically desirable, previous experiments have shown these blisters to have no observed effect on the mechanical loss\cite{Penn2019mechanical}.
These are the largest  \GAG coatings for which mechanical loss measurements have been performed.  As of this writing,  20~\unit{cm} \GAG mirrors are now in production and the first measurements at this larger size have recently been published~\cite{gretarsson2026nonuniform}.

\section{Experimental Methods}\label{sec:Experiment}

Following the method detailed in previous experiments\cite{Harry2002thermal, Penn2003mechanical}, we have measured the quality factor, $Q_i$ at several normal modes with resonant frequencies, $f_i$.  The amplitude of a  mode brought to resonance and allowed to freely ringdown is a damped sinusoid with decay time, $\tau_i$, and phase angle, $\delta$.
\begin{equation} \label{eqnRingdown}
    A_i(t) = A_0 e^{-t/\tau_i} \cos\left(2 \pi f_i t + \delta \right)
\end{equation}
\noindent With the quality factor given by $Q_i =\pi f_i \tau_i$.

The GeNS system used in this experiment will, before each measurement, balances the wafer sample 
on the face of a firmly-anchored lens to provide a single point of support.  For modes where the support point is at a node of the oscillation, these systems have demonstrated very low frictional loss.  However, if the support point is not at a node, the contact frictional losses will obscure the loss in low dissipation materials.

\begin{figure}[htbp]
    \centering
    \graphicspath{ {./Figures/} }
    \includegraphics[width=8.6cm]{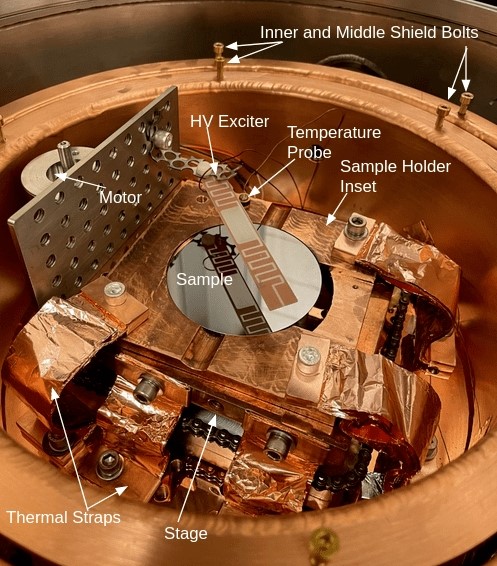}
    \caption{The interior of the Cryo-GeNS chamber. The sample is a $10~\unit{cm}$ diameter bare silicon substrate.}
    \label{fig:Cryo_GeNS}
\end{figure}

The Cryo-GeNS apparatus (Fig.~\ref{fig:Cryo_GeNS}) is housed inside a cryostat chamber, developed in cooperation with Cryomech (now Bluefors) \cite{Cryomech}.
The inner and middle shields thermally isolate the apparatus and allow it to cool down to $<12~\unit{K}$.
Except during data collection runs, the sample is held by the outer edge in a sample holder, a copper plate that is kept at equilibrium temperatures through flexible copper straps anchored to the main cryostat baseplate. The sample holder is attached to a
custom-designed, vertical stage developed to operate in cryogenic vacuum.  The stage was designed and build in cooperation with Cryomech and is driven by a motor from Empire Magnetics~\cite{EmpireMagnetics}.
When thermal equilibrium is achieved, the sample holder is lowered and the sample is  balanced on the nodal support.  The precision of this repeatable mounting process is such that the optical system does not require realignment between runs. 
After the sample is suspended, its resonant modes are excited by a high-voltage comb exciter.

To ring-up the resonant modes, the exciter is driven by a  high voltage, white noise spectrum spanning the range of the resonant modes to be observed.  The resonant motion is detected using an optical lever, which consists of a HeNe laser injected into the cryostat chamber and  onto the upper surface of the sample, and the reflected beam directed out of the chamber and onto a quadrant photodetector (QPD). The QPD outputs are amplified and converted to X, Y, and SUM values. The optical lever arm was 2~\unit{m}. The laser power at the QPD was about $100~\unit{\mu W}$.
These parameters yielded a conversion factor of the disk's angular motion into the normalized QPD signal of $\sim1.6\times10^4~\unit{rad^{-1}}$, which is comparable to previous experiments  \cite{Vajente2017GeNS}.

During each measurement, the temperature is monitored by two temperature probes attached to the sample holder and the floor of the chamber, close to the nodal suspension base. The chamber pressure is below 
$10^{-2}~\unit{mPa}$ 
for $T<63~\unit{K}$. Above that point the pressure slowly increases to 
$3~\unit{mPa}$ 
at room temperature.
Therefore, viscous gas damping is well below the sample loss across the full temperature range, as in previous studies \cite{Zendri2008, Cesarini2009Gentle, Vajente2017GeNS}.

The MultiQ~\cite{Penn_MultiQ} application was used to perform a multimodal acquisition of the sample ringdown data. Written in LabView, MultiQ simultaneously measures all of the sample modes of interest by first driving them all to resonance and then recording sample motion at a high data rate to capture all the mode ringdowns.   The application drives the exciter using  white voltage noise spanning the frequency range of the modes to be measured.  MultiQ  then records the X, Y, and SUM data channels at a sampling rate for which the  Nyquist frequency is well above the highest mode frequency. The data is continuously saved to disk as it is being collected.   For mode frequencies of $f=390 - 8100~\unit{Hz}$, the data was collected at $22~\unit{kHz}$. 
A multimodal readout is well designed for cryogenic measurements since it ensured that all of the modes recorded for a given run are taken with identical conditions.

The X and Y data, normalized by SUM, are then heterodyned to extract the time series for each mode from the multimodal data.  
The  data is modulated by a reference signal, $f_r$, close to the mode frequency, $f_i$, with $f_h = \left|f_i - f_r\right| < 1~\unit{Hz}$.  The data is then low-pass filtered at $2 f_h$ and resampled at sample frequency, $f_s = 10\,f_h$.  For samples with high levels of symmetry, the frequency separation of degenerate modes is small but easily distinguishable over the standard run duration of $2 \tau$.  The asymmetry of the samples used in this experiment allowed the degenerate modes to be easily distinguished.  
The resulting time series data for each mode is then fit to a damped sinusoid using the QView~\cite{Penn_QView} application which outputs the time constant $\tau_i$ and quality factor $Q_i$.

\section{Modeling the Sample Loss}\label{sec:LossModel}

We modeled the dissipation in the measured samples using four loss mechanisms:  Thermoelastic loss $\left(\phi_{\mathrm{TE}}\right)$,  Phonon-phonon scattering loss $\left(\phi_{\mathrm{PP}}\right)$, Friction $\left(\phi_{\mathrm{F}}\right)$, and Elastic loss $\left(\phi_{\mathrm{E}}\right)$.  To extract the elastic loss of the substrate and coating, we model and subtract the other loss mechanisms.  We will discuss each of these loss terms below.

\subsection{Thermoelastic loss}\label{sec:TELoss}

In any elastically deformed body, the strain field drives a change in the internal energy such that compressed regions are heated and extended regions cooled for positive coefficients of thermal expansion (CTE). (A negative CTE inverts that coupling.) These temperature gradients drive irreversible heat flow (dissipation) known as thermoelastic (TE) damping \cite{Zener1937}.  This damping is maximized when the heat travel time is equal to the mode period.  For the silicon wafer substrate, the TE loss dominates all other loss mechanisms above 150~\unit{K}. For simple modal geometries with uniform materials, such as the transverse modes of a fiber, the TE loss can be modeled analytically.  
For example, the TE loss of a uniform disk, is given by \cite{Cagnoli2018},
\begin{equation}\label{eqn:SubstrateTED}
    \phi_{\mathrm{TE},i} = \mathcal{D}_i\frac{(3\alpha)^2KT}{\rho C}\frac{\omega_i\omega_{\mathrm{peak}}}{\omega_i^2+\omega^2_{\mathrm{peak}}},
\end{equation}
where 
$\alpha$ is the linear thermal expansion, 
$K$ is the bulk modulus, $T$ is the disk temperature, 
$C$ is the specific heat capacity, $\rho$ is the density, 
$\omega_i = 2\pi f_i$ is the mode angular frequency, $\mathcal{D}_i$ is the geometric factor determined by the mode shape, and $\omega_{\mathrm{peak}}$ is the Debye peak given by 
\begin{equation}\label{eqn:UncoatedTEDOmega}
    \omega_{\mathrm{peak}} = \frac{\kappa}{\rho C}\Big(\frac{\pi}{h_{\mathrm{s}}}\Big)^2.
\end{equation}
In the above expression, 
$\kappa$ is the thermal conductivity and 
$h_{\mathrm{s}}$ is the thickness of the disk.

 We calculated the thermoelastic loss using a finite element model, developed in COMSOL, because of the  material asymmetry and geometric complexity of the nodally suspended sample. The silicon substrate and GaAs/AlGaAs coating were model as crystalline materials with asymmetric elastic tensors. The multilayer coating, which consisted of 38.5 doublets, was approximated as a single effective layer using the method described by Fejer\cite{Fejer2021EffectiveMedium}. The geometric asymmetry of the coating resulted in the effective parameters being decomposed into normal and transverse components with respect to the surface normal. The interface between the substrate and coating assumed full contact between the materials with no dissipation or resistance to heat flow.  The model required that the material parameters ($\alpha$, $\kappa$, $\rho$, $C$) and the elastic constants $\left(C_{11}, C_{12}, C_{44}\right)$ for Si, \GA, and \AG  be modeled as smooth functions of temperature from 12--300~\unit{K}.  The temperature-dependent functions for these parameters are well established for Si and GaAs, but were unavailable for \AG.  Using the room-temperature parameters for \AG, we approximated the temperature-dependent functions by scaling the \GA functions by the ratio of \AG-to-\GA parameter values at room temperature. As in our experiment, the COMSOL model had the sample balanced at its geometric center on a lens that maintained the constant temperature of the cryostat.  The TE loss was calculated for all modes at all measured temperatures.  


\subsection{Phonon-phonon scattering loss}\label{sec:AkhiezerLoss}
In the model, a material's thermal energy can be characterized as a phonon distribution. Similarly, the vibrational energy of a sample at resonance can also be described by a phonon distribution. Akhiezer first determined that vibrational phonons can scatter off the thermal phonons and dissipate energy from the resonant mode. This ``Akhiezer Loss'' is a low magnitude loss that has been observed in silicon flexures\cite{Nawrodt2008Si, Nawrodt2013Si, rodriguez_direct_2019}. Adopting the formulation used by Rodriguez\cite{rodriguez_direct_2019}, the phonon-phonon loss, $\phi_{\mathrm{PP}}$, is goven by
\begin{equation}\label{eqn:Akhiezer}
\phi_{\mathrm{PP}} = \frac{  \gamma_{\mathrm {eff}}^2 \, \omega \, T \, \kappa}{\rho \,v^2 \, v_{\mathrm{D}}^2}
\end{equation}
where $\gamma_{\mathrm{eff}}$ is a mode-dependent geometric factor, $v$ is the acoustic velocity, and  the Debye velocity, $v_{\mathrm{D}}$, is given by
\begin{equation}
    v_{\text{D}}^{-3} = \frac{1}{3}\left[ v_{\text{L}}^{-3}  + v_{\text{T1}}^{-3}  + v_{\text{T2}}^{-3}       \right]
\end{equation}
and $v_{\text{L}}$, $v_{\text{T1}}$, and $v_{\text{T2}}$ are the  acoustic velocities along the longitudinal and 2 transverse directions  respectively.

The distinct shape of the Akhiezer loss in silicon arises primarily from the thermal conductivity, with a sharp rise at low $T$ and a long tail to high $T$.   The thermal conductivity peak temperature, width, and magnitude will depend on the doping concentrations and on the sample geometry.  


\subsection{Contact friction}\label{sec:Friction}

The flat of the silicon wafer substrate shifted the center of mass and central nodal point away from the support point, undermining the primary benefit of the GeNS system, that the support point is a nodal point.  The sample's vibration on the support point produces contact friction that is large in comparison to the sample's elastic loss.  Considerable effort was devoted to modeling this contact friction using COMSOL, with the intent of subtracting this excess loss.  However, those efforts to model these extremely small levels of friction were ultimately not successful.  As a result, the modes with  high levels of contact friction were excluded from the final analysis step to determine $\phi_{\mathrm{E}}$.  To distinguish the modes with excess friction, we extracted from our model the maximum deviation at the support point for each mode.  

\section{Analysis}\label{sec:Analysis}
\subsection{Loss in silicon wafers}

The mechanical loss measurements were performed on an uncoated and a \GAG-coated silicon wafer for 8 modes with frequencies up to $8.1~\unit{kHz}$ and temperatures from 12 -- 300 K.  
For modes that are susceptible to contact friction, we observe an increased variance in the measured results, as is reflected in the large error bars.

\begin{figure}[hbt]
    \centering
    \graphicspath{ {./Figures/} }
    \includegraphics[width=8.6cm]{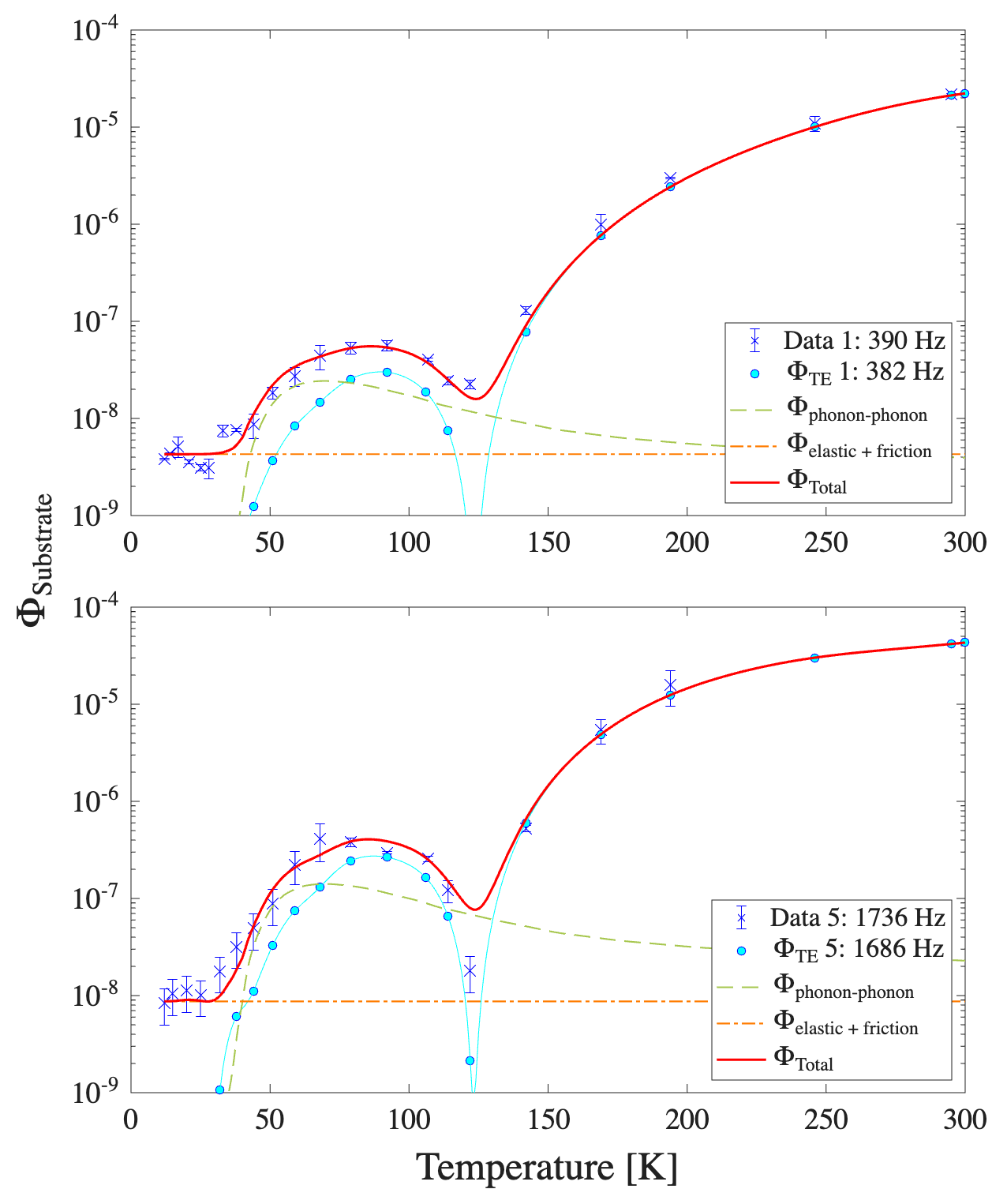}
    \caption{The losses in the silicon wafer for the low friction modes including elastic, thermoelastic, phonon-phonon, and frictional losses. }
    \label{fig:SubstrateMode1}
\end{figure}

\begin{figure}[hbt]
    \centering
    \graphicspath{ {./Figures/} }
    \includegraphics[width=8.6cm]{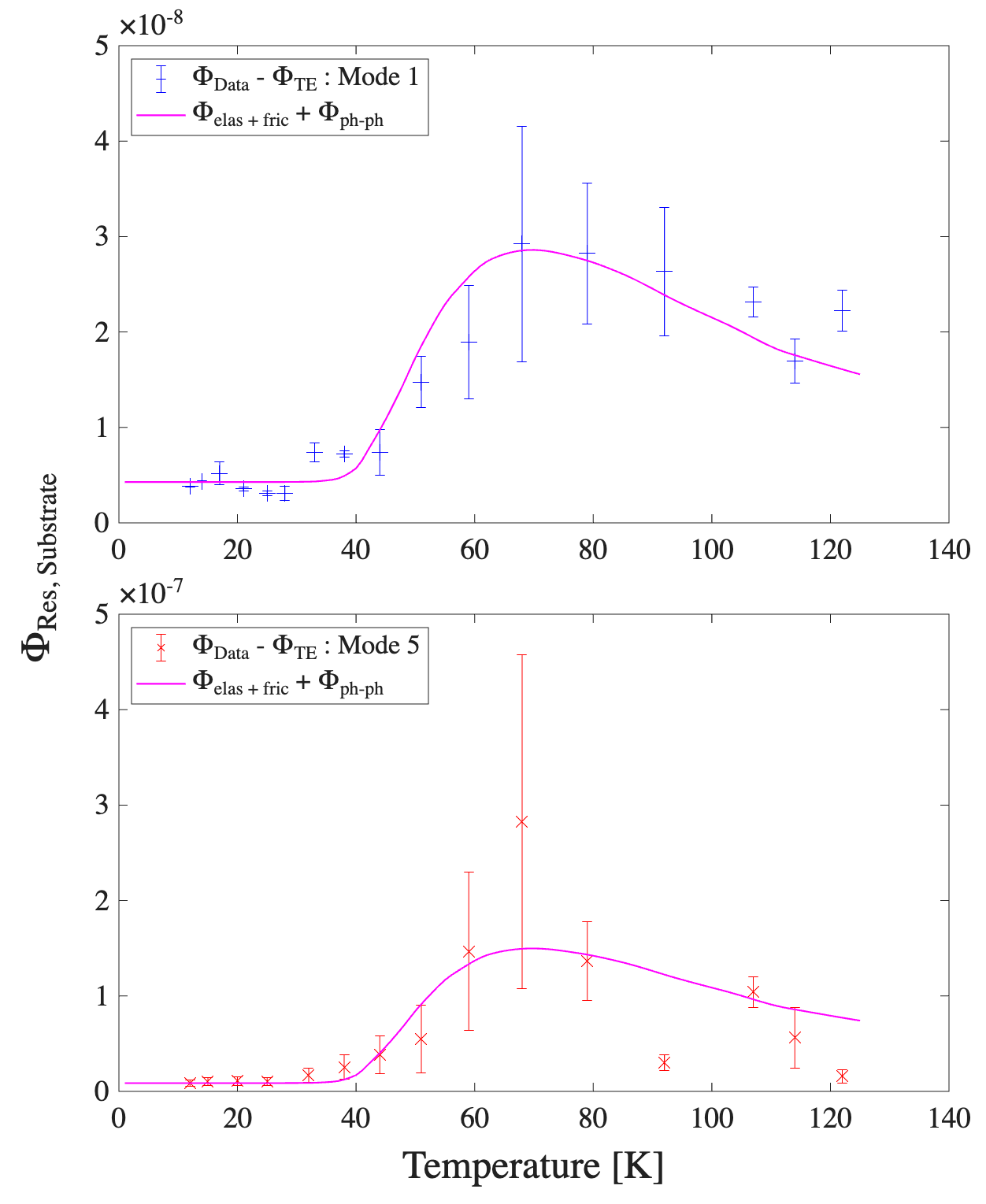}
    \caption{The residual loss in the silicon wafer for the low friction modes including elastic, phonon-phonon, and frictional losses. }
    \label{fig:SubstrateResidualMode1}
\end{figure}

The loss terms discussed in this analysis will be identified by 2 subscripts: the first, upper-case subscript indicates the loss type (\textbf{M}easured, \textbf{T}hermo\textbf{E}lastic, \textbf{P}honon-\textbf{P}honon, \textbf{E}lastic, \textbf{F}riction, \textbf{R}esidual) and the second, lower-case subscript denotes the loss medium (\textbf{S}ubstrate, \textbf{C}oating, or \textbf{C}oated \textbf{S}ubstrate).

As is shown for Mode 1 in Figure~\ref{fig:SubstrateMode1}, the measured loss of the silicon substrate, $\phi_{\mathrm{M,s}}$, has 4 components: 
\begin{equation}\label{eqnSubstrateLoss}
    \phi_{\mathrm{M,s}} = \phi_{\mathrm{TE,s}} + \phi_{\mathrm{PP,s}} + \phi_{\mathrm{E,s}} + \phi_{\mathrm{F,s}}
\end{equation}
As described in Section~\ref{sec:TELoss}, the thermoelastic loss, $\phi_{\mathrm{TE}}$ was modeled using COMSOL. 
Subtracting the thermoelastic loss from the measured loss, we obtain the residual loss:
\begin{equation}\label{eqnResidualLoss}
    \phi_{\mathrm{R,s}} = \phi_{\mathrm{M,s}} - \phi_{\mathrm{TE,s}} =  \phi_{\mathrm{PP,s}} +\phi_{\mathrm{E,s}} + \phi_{\mathrm{F,s}} 
\end{equation}
as is shown in Figure~\ref{fig:SubstrateResidualMode1}. A compilation of the loss plots for all 8 modes is given in the appendix.

For those modes with lower uncertainty, the shape of the residual loss was flat for $T < 25~\unit{K}$.
For $T > 35~\unit{K}$, the loss displayed the characteristic shape of phonon-phonon scattering loss.  Therefore, we fit the data to the form 
\begin{equation}
    \phi_{\mathrm{R,s}}  =  \phi_{\mathrm{PP,s}} + \left(\phi_{\mathrm{E,s}} + \phi_{\mathrm{F,s}}\right) = \phi_{\mathrm{PP,s}} + \phi_{\mathrm{constant}} 
\end{equation}
The form of the phonon-phonon loss is largely defined by the thermal conductivity, $\kappa_{\mathrm{Si}}$.  We modeled $\kappa_{\mathrm{Si}}$ by transforming the data measured for bulk silicon~\cite{Glassbrenner1964} by shifting the peak centroid, width, and amplitude to match the observed loss.  The constant and phonon-phonon loss functions are not orthogonal. Friction increases the phonon energy and the observed phonon-phonon loss. However, for the low friction modes, $\kappa_{\mathrm{Si}}$ and $\phi_{\mathrm{PP,s}}$ were in agreement with other reported results~\cite{Nawrodt2013Si}.  

\begin{figure}[hbt]
    \centering
    \graphicspath{ {./Figures/} }
    \includegraphics[width=8.6cm]{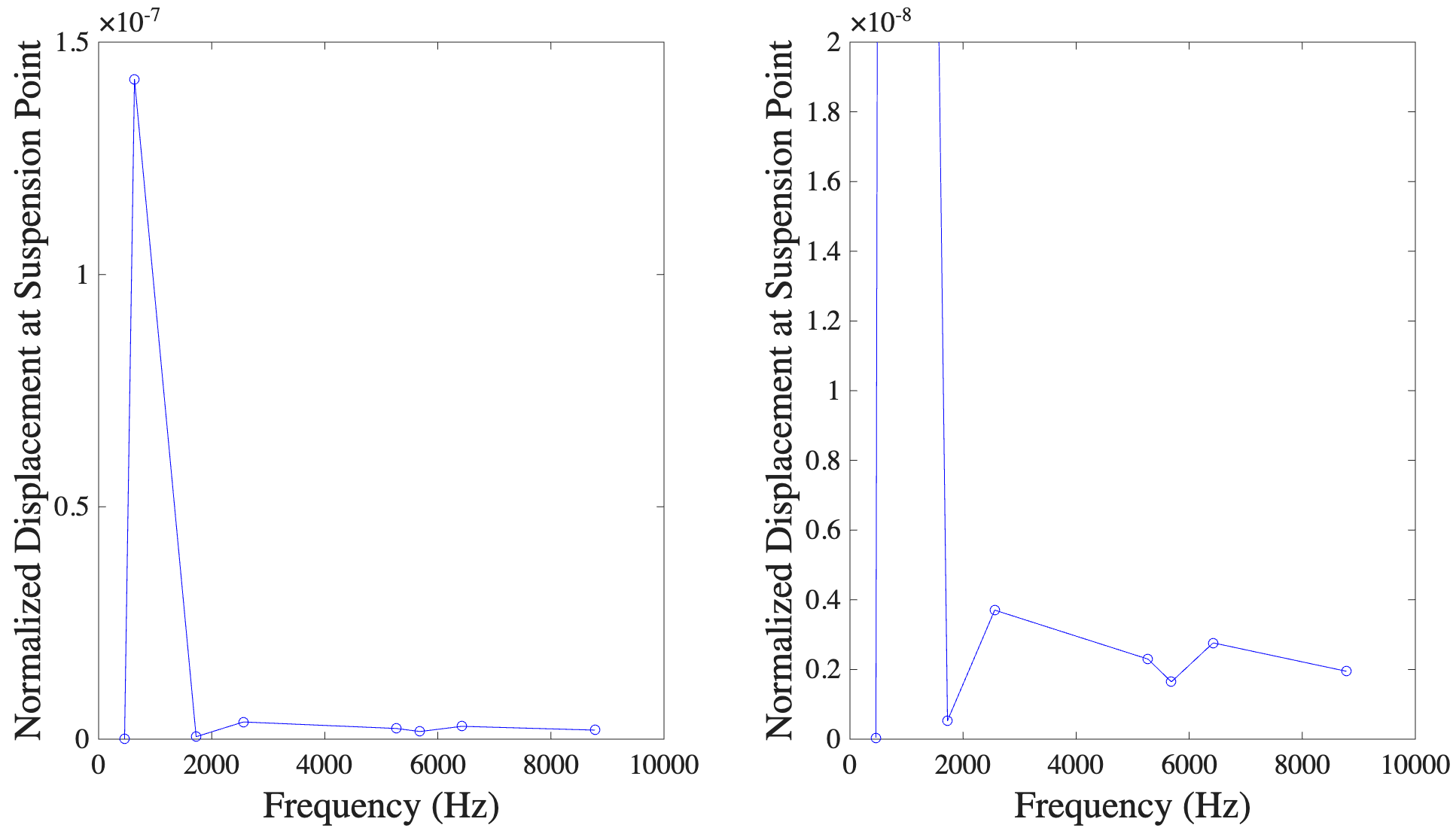}
    \caption{The z displacement at the suspension point for the uncoated silicon wafer. There is a monotonic mapping of higher displacement to higher frictional loss.}
    \label{fig:SubstrateDisplacement}
\end{figure}

\begin{figure}[hbt]
    \centering
    \graphicspath{ {./Figures/} }
    \includegraphics[width=8.6cm]{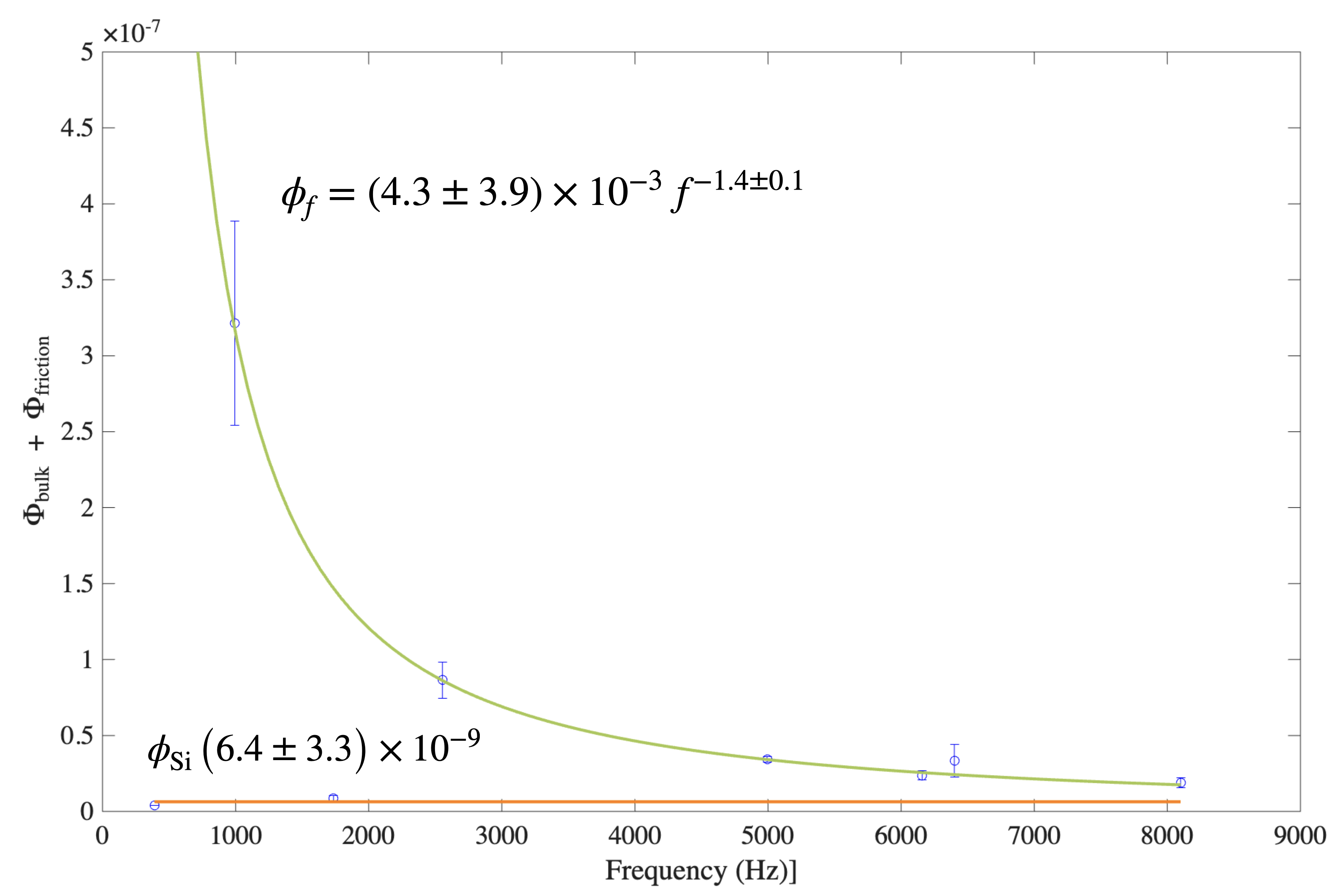}
    \caption{The combined elastic and friction loss in the silicon wafer are separated into the modes with minimal friction, $\phi_{\mathrm{E,s}} = \phi_{\mathrm{Si}} = \left(6.4\pm 3.3\right) \times 10^{-9}$, and the modes dominated by frictional losses, with a power law decrease in loss as the displacement at the support point declines. }
    \label{fig:SubstrateFriction}
\end{figure}

The constant loss, $\phi_{\mathrm{constant}}$
displayed a high level of variability 
that was correlated with the displacement at the support point as derived from the finite element model and shown in Figures~\ref{fig:SubstrateDisplacement} and \ref{fig:SubstrateFriction}.  We assume that the excess loss seen in modes with large displacement at the support point to arise from contact friction between the substrate and the supporting lens.  
In Figure~\ref{fig:SubstrateFriction} we perform an empirical fit of this frictional loss to a decreasing power law, $\phi \propto f^{-1.4 \pm 0.1}$, as the displacement at the support point declines. Assuming that the friction is negligible in the low loss modes, the data predicts an elastic loss of $\phi_{\mathrm{E,s}} = \left(6.4 \pm 3.3\right) \times 10^{-9}$ which lies between the cryogenic elastic loss results measured by Nawrodt~\cite{Nawrodt2008Si,Nawrodt2013Si}, who states that samples of this geometry are surface loss limited. 

Using only those modes with no contact friction, we find that the best fit of $\phi_{\mathrm{PP,s}}$ yields a $\kappa_{\mathrm{Si}}$ whose centroid
at $T=70~\unit{K}$ for the 500 $\mu$m thick wafer agrees well with the measured conductivity in a 775 $\mu$m thick silicon fiber.~\cite{Toland_2025} 

\subsection{Loss in \GAG-coated silicon wafers}

The process for analyzing the \GAG-coated silicon wafers is quite similar to the analysis of uncoated samples.  Therefore, in our description we will focus on the points of difference.  First, the thermoelastic loss was modeled for the entire coated sample, rather than simply the added coating, in order to include thermoelastic losses across the coating-substrate interface.  However the residual losses for the coating and substrate were analyzed separately.  Thus the measured loss for the coated sample, $\phi_{\mathrm{M,cs}}$ has the form:
\begin{equation}\label{eqnCoatedLoss}
\begin{split}
    \phi_{\mathrm{M,cs}} = \phi_{\mathrm{TE,cs}} & + \left(\phi_{\mathrm{PP,s}} + D_{\mathrm{s}}\, \phi_{\mathrm{E,s}}  + \phi_{\mathrm{F,s}}\right) \\ & + \left(\phi_{\mathrm{PP,c}} + D_{\mathrm{c}}\,\phi_{\mathrm{E,c}}  + \phi_{\mathrm{F,c}}\right) \\
\end{split}
\end{equation}
where the dilution factors, $D_{\mathrm{c}} = E_{\mathrm{c}}/E_{\mathrm{cs}}$ and $D_{\mathrm{s}} = E_{\mathrm{s}}/E_{\mathrm{cs}} \approxeq 1$,  are the fractional energy in the coating and substrate. 

We then define the residual loss for the coating as: 
\begin{equation}\label{eqnCoatedResidual}
    \phi_{\mathrm{R,c}} = \phi_{\mathrm{M,cs}} - \phi_{\mathrm{TE,cs}} -  \phi_{\mathrm{R,s}} = \phi_{\mathrm{PP,c}}  +  \phi_{\mathrm{F,c}} + D_{\mathrm{c}}\,\phi_{\mathrm{E,c}} 
\end{equation}
where we assume that $D_{\mathrm{s}} = 1$.  We note that $\phi_{\mathrm{F,c}}$ represents the increased friction that was observed in the coated sample, rather than any direct external friction with the coating surface.  Also,  as was noted in the substrate analysis, friction  generates additional phonon energy and increases the phonon-phonon loss.  However, since we will ultimately exclude those modes with excess frictional loss, our final result will not be effected by that process.  

As with the substrate residual loss, we fit the coating residual loss as 
\begin{equation}
    \phi_{\mathrm{R,c}}  =  \phi_{\mathrm{PP,c}} + \phi_{\mathrm{F,c}} + D_{\mathrm{c}}\,\phi_{\mathrm{E,c}} = \phi_{\mathrm{PP,c}} + \phi_{\mathrm{constant}} 
\end{equation}
except that we use the properties for the  \GAG effective layer for $\phi_{\mathrm{PP,c}}$. The total loss and residual loss for the low friction modes are shown in Figures~\ref{fig:CoatedMode1} and \ref{fig:CoatedResidualMode1}. The plots for all modes are shown in the appendix.

\begin{figure}[hbt]
    \centering
    \graphicspath{ {./Figures/} }
    \includegraphics[width=8.6cm]{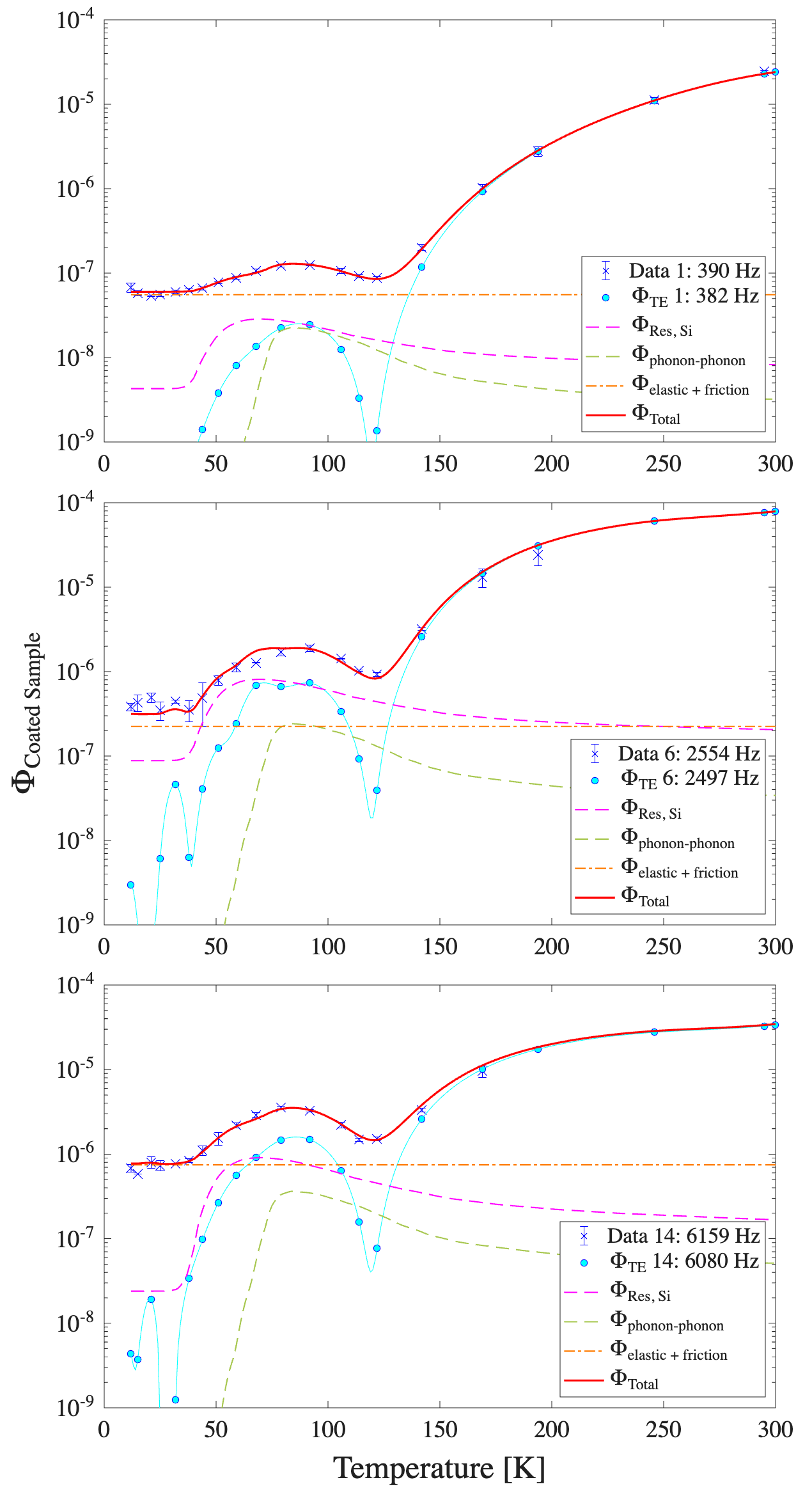}
    \caption{The losses in the \GAG-coated silicon wafer for the low friction modes including elastic, thermoelastic, phonon-phonon, and frictional losses. }
    \label{fig:CoatedMode1}
\end{figure}

\begin{figure}[hbt]
    \centering
    \graphicspath{ {./Figures/} }
    \includegraphics[width=8.6cm]{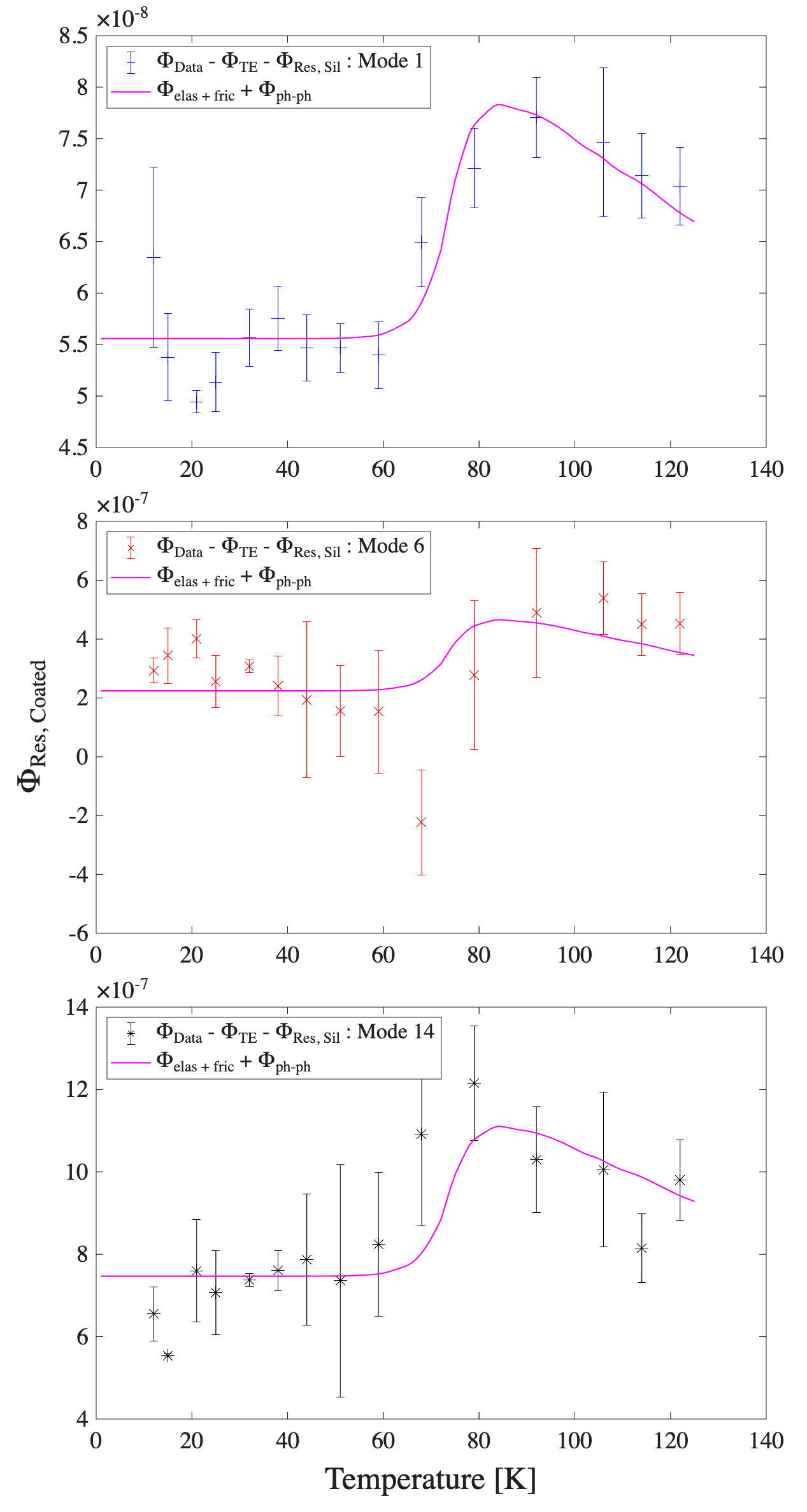}
    \caption{The residual loss in the  \GAG-coated silicon wafer for  the low friction modes including elastic, phonon-phonon, and frictional losses. }
    \label{fig:CoatedResidualMode1}
\end{figure}

\begin{figure}[hbt]
    \centering
    \graphicspath{ {./Figures/} }
    \includegraphics[width=8.6cm]{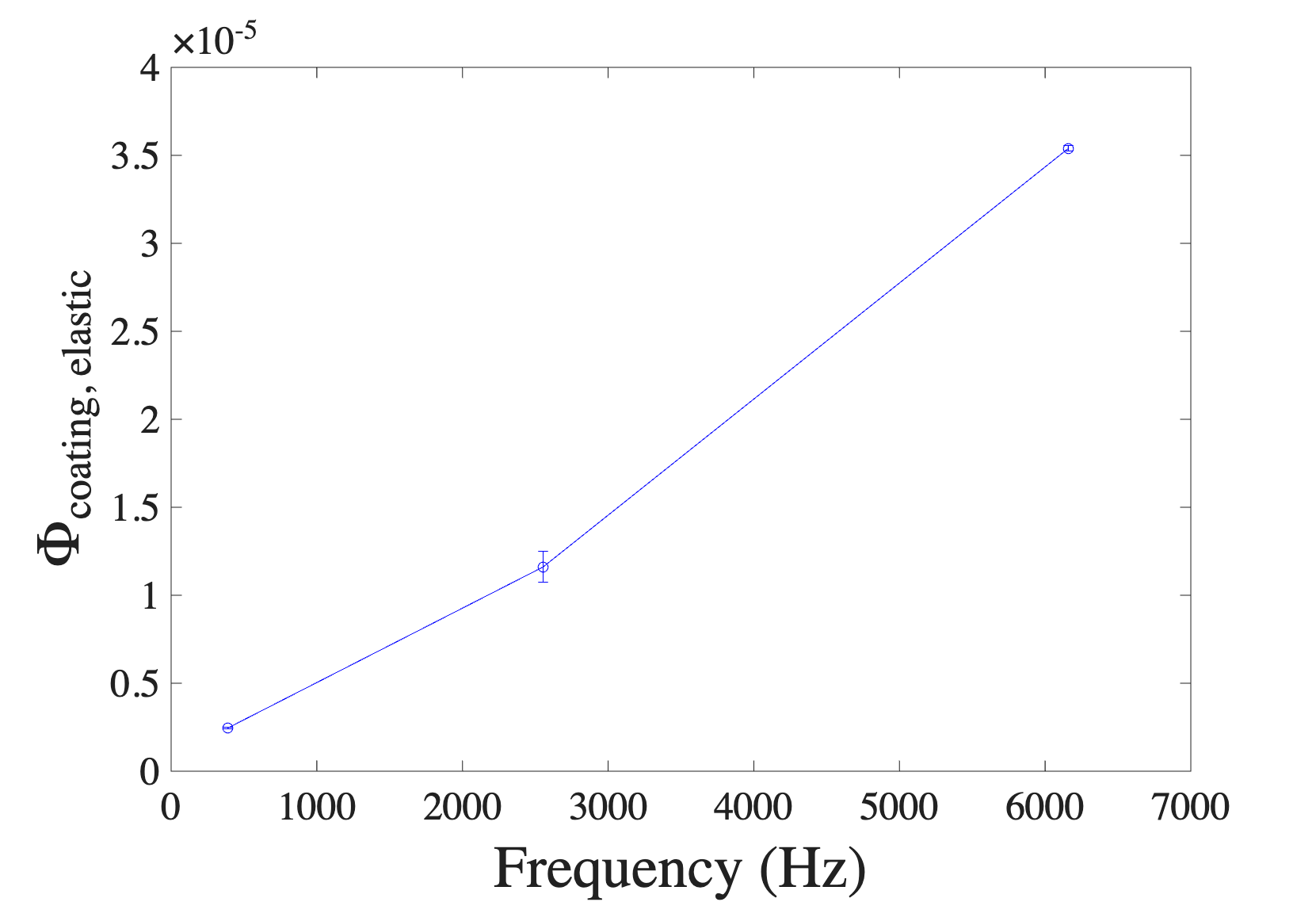}
    \caption{The effective cryogenic elastic loss of \GAG coating material.}
    \label{fig:CryoGAGLoss}
\end{figure}

Finally, if we restrict ourselves to the modes where the contact displacement is low and the frictional loss is assumed negligible, we can calculate the elastic loss in the \GAG coating material, $\phi_{\mathrm{E,c}}$ using
\begin{equation}\label{eqnCoatingLoss}
    \phi_{\mathrm{E,c}} = \frac{\phi_{\mathrm{constant}}}{D_{\mathrm{c}}}
\end{equation}



\section{Results}\label{sec:Results}
The cryogenic elastic loss of \GAG is shown in Figure~\ref{fig:CryoGAGLoss}.  Exclude modes with high contact friction, the elastic loss (internal friction) is observed as the dominant loss mechanism in the range of $\approx 8$--$25~\unit{K}$. At higher temperatures, other loss mechanisms dominate and mask any variation in the elastic loss.  Thus, only in this temperature range can we determine the elastic loss of \GAG, and even these measurements should be considered as an upper limit. 

Indirect measurements of the coating elastic loss derived from optical cavity measurements indicate that $\phi_{\mathrm{E,c}}$ at $17~\unit{K}$ ranges from $6\times 10^{-6}$~\cite{Cole2012} to $4 \times 10^{-5}$~\cite{pagano2026}.  Recent results by Lee~\cite{Lee2026PRL} on the currently lowest noise laser system predict $\phi_{\mathrm{E,c}}$ to be $2.3\times 10^{-5}$. 
A cryogenic optomechanical cantilever experiment~\cite{Cole2012} predicted a $\phi_{\mathrm{E,c}} \approx4.5\times10^{-6}$.
Our measured results for $\phi_{\mathrm{E,c}}$, ranging from $2.5\times 10^{-6}$ to $3.3\times 10^{-5}$ are in general agreement with these results and are comparable to or lower than the room temperature results.~\cite{Cole2013, Penn2019mechanical}.

In the range of $\approx 8$--$25~\unit{K}$ we can use the elastic loss to predict the coating thermal noise. 
The Einstein Telescope Low Frequency detector (ET-LF), a next-generation GW detector with a cryogenic design, is planned to operate at 10 K.
The ET-LF design goal lists the total CTN for the arm cavity as $\approx 3.6\times 10^{-21}\,\mathrm{m/\sqrt{Hz}}$ and absorption in the HR coating $a_{\mathrm{HR}} \le 5\unit{ppm}$.  As reported by Craig~\cite{Craig2019ET}, the currently proposed coating is a multimaterial design where the outer 2 doublets of \silica/\tantala have low optical absorption and high elastic loss while the inner 10 doublets of \silica:\hafnia/\asi have low elastic loss and higher absorption.  The resulting HR coating on the end test mass (ETM) has an  $a_{\mathrm{HR}} = 3.4~\unit{ppm}$ and CTN of $1.9\times 10^{-21}\,\mathrm{m/\sqrt{Hz}}$. The CTN of the intial test mass (ITM) is $1.6\times 10^{-21}\,\mathrm{m/\sqrt{Hz}}$ for a total CTN of $3.5\times 10^{-21}\,\mathrm{m/\sqrt{Hz}}$.

In cryogenic fixed cavity experiments, \GAG coatings have repeatedly demonstrated the lowest CTN.  Using our measured results for the cryogenic elastic loss, we can use Eqn~\ref{eqn:ctn} to calculate the expected CTN for ET-LF HR coatings.  

\begin{equation}
    x\left(f\right) = \sqrt{\frac{2 k_{\mathrm{B}}T}{\pi^2 f} \frac{d}{w^2} \frac{\phi}{Y_{\mathrm{sub}}} \left(\frac{Y_{\mathrm{coat}}}{Y_{\mathrm{sub}}} + \frac{Y_{\mathrm{sub}}}{Y_{\mathrm{coat}}}\right)}
    \label{eqn:ctn}
\end{equation}

The ET-LF requirements assume a temperature, frequency, and beam radius of  $T=10\, \unit{K}$, $f=10\, \unit{Hz}$, and $w=9\, \unit{cm}$ respectively.  \GAG HR coatings at $\lambda = 1550\, \unit{nm}$ have a thickness $d=12\, \unit{\mu m}$.  The substrate and coating Young's moduli are $Y_{\mathrm{sub}}=188\, \unit{GPa}$    $Y_{\mathrm{coat}}=100\, \unit{GPa}$.  

\begin{equation}
    x_{\mathrm{G/A}}\left(10\,\unit{Hz}\right) = 2.3\times 10^{-19}\,\sqrt{\phi_{\mathrm{E,c}} } \,\,\,\mathrm{m/\sqrt{Hz}}
    \label{eqn:ETLFCTN}
\end{equation}


At $f = 390\,\unit{Hz}$, our lowest frequency, $\phi_{\mathrm{E,c}} = 2.5 \times 10^{-6}$, which predict a CTN for the HR coating of $3.6\times 10^{-22} \,\,\mathrm{m/\sqrt{Hz}}$.  Using the mean of our results, $\phi_{\mathrm{E,c}} = 1.6 \times 10^{-5}$  predict a CTN for the HR coating of $9.2\times 10^{-22} \,\,\mathrm{m/\sqrt{Hz}}$.  These results are respectively a factor $5\times$ and $2 \times$ lower than the current ET-LF coating candidate~\cite{Craig2019ET}. Thus adopting \GAG coatings would provide a significant sensitivity improvement for the ET-LF detector.

It should be noted that at the time of this writing there are active investigations of generation-recombination noise in room temperature \GAG coatings~\cite{wu2025}.  However, the predicted temperature dependence indicates that this noise source is negligible at these cryogenic temperatures.

\begin{figure*}
\centering
\includegraphics[width=1\textwidth]{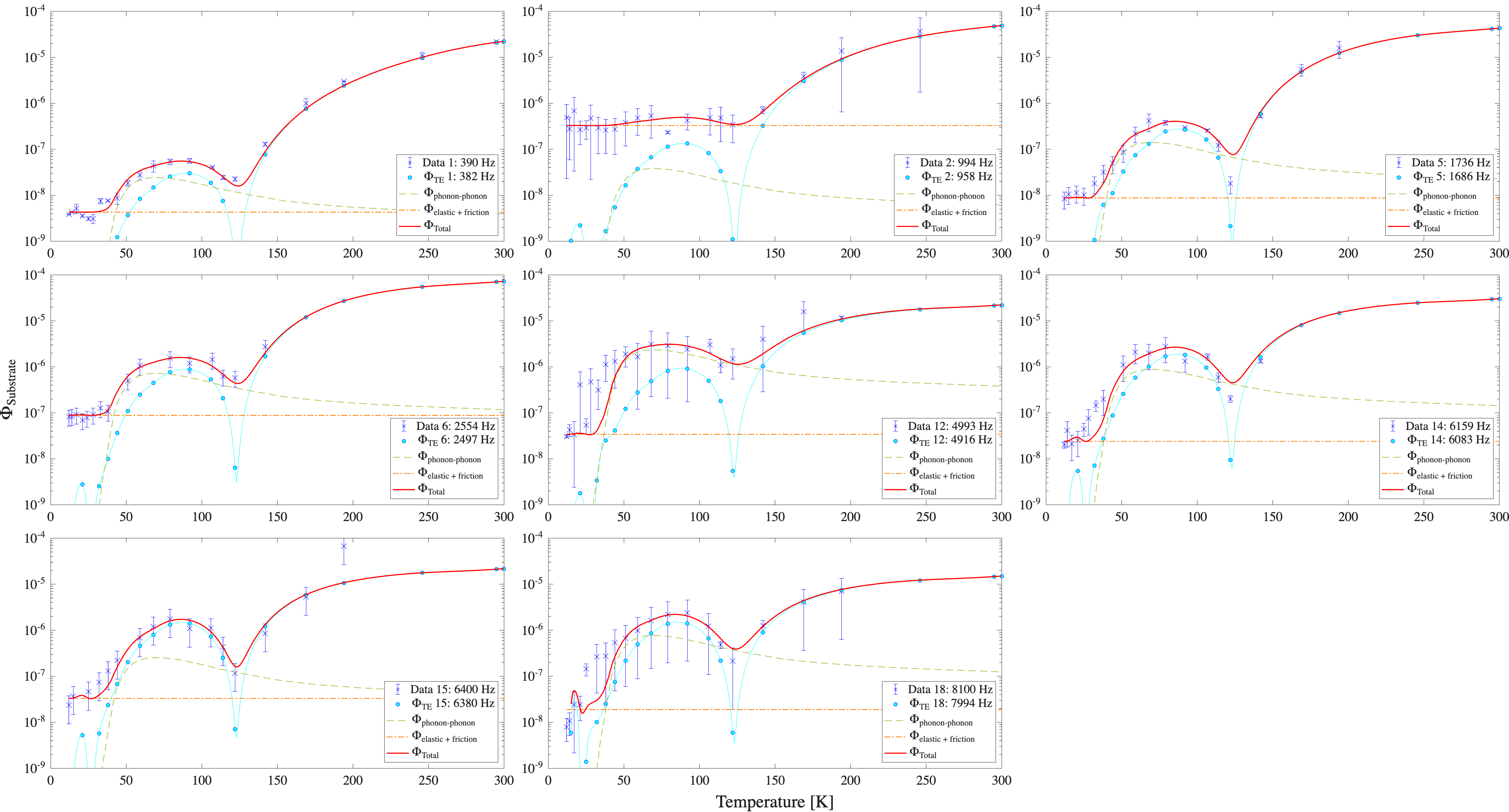}
\caption{The losses for all modes in the silicon wafer, including elastic, thermoelastic, phonon-phonon, and frictional losses.}
\label{fig:SiliconLossAll}
\end{figure*}

\begin{figure*}
\centering
\includegraphics[width=1\textwidth]{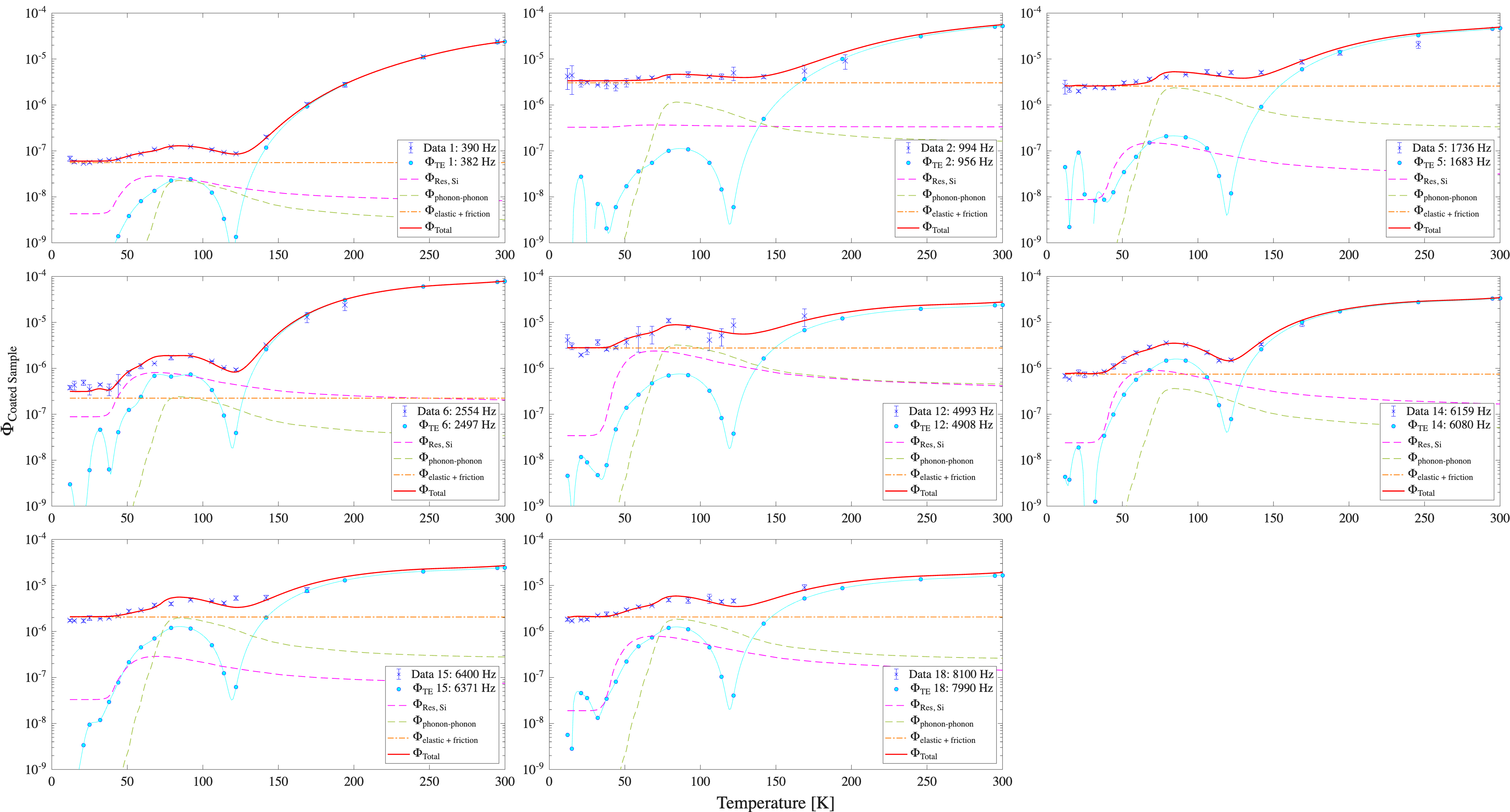}
\caption{The losses for all modes in the \GAG-coated silicon wafer, including elastic, thermoelastic, phonon-phonon, and frictional losses.}
\label{fig:AGCoatLossAll}
\end{figure*}

\section{Conclusion}\label{sec:Conclusion}

We have performed the first direct measurement of the elastic loss in \GAG coatings at cryogenic temperatures.  The results are in agreement or slightly below the loss values inferred from cryogenic optical cavity measurements.  \GAG coatings have already proven their superior low noise characteristics in the cryogenic optical cavities used in precision time-frequency measurements~\cite{Lee2026PRL}.  Our results indicate that if \GAG coatings were used in the cryogenic Einstein Telescope - Low Frequency GW detector it would result in at least a $2\times$ reduction in coating thermal noise.  Room temperature measurements of the \GAG CTN indicate a reduction in CTN of 3--10$\times$ compared to the current Advanced LIGO coating.  Thus the same coating technology could be employed for room temperature and cryogenic gravitational wave detectors and in so doing provide the lowest coating thermal noise for both designs. 

These measurements were performed on the  Cryo-GeNS system, developed by the Syracuse GW research team to perform coating mechanical loss measurements from 12 -- $300\,\unit{K}$.  Upgrades to this system are currently underway with the intent of measuring the elastic loss in \GAG with greater precision and over a wider temperature range.

\begin{acknowledgments}
We would like to thank Professor Martin Fejer for many insightful discussions. 
We also thank the LSC Optics Working Group for valuable feedback, and constructive criticism.

The development of the Cryo-GeNS system would not have been possible without the vital assistance of the Cryomech team, especially Brent Zerkle, Tim Hanrahan, and Kayleigh Byrns,
and the expertise and guidance of the Syracuse University Physics Machine Shop machinists, especially Phil Arnold. 

This work was supported with funding from the National Science Foundation Grants No. PHY-1920023, PHY-2011723,  PHY-2207640, PHY-2208079, PHY-2309296, PHY-2409601, and PHY-2513058. 
This paper is assigned LIGO Document number LIGO-P2600350.
\end{acknowledgments}




\bibliography{cryoloss}

\end{document}